\documentclass[11pt,preprint]{aastex}

\shorttitle{Far Ultraviolet Spectral Images of the Cygnus Loop}
\shortauthors{Seon et al.}

\begin{document}

\title{{\em SPEAR} Far Ultraviolet Spectral Images of the Cygnus Loop}%

\author{Kwang-Il Seon\altaffilmark{1},
Wonyong Han\altaffilmark{1},
Uk-Won Nam\altaffilmark{1},
Jang-Hyun Park\altaffilmark{1},\\
Jerry Edelstein\altaffilmark{2},
Eric J. Korpela\altaffilmark{2},
Ravi Sankrit\altaffilmark{3}, \\
Kyoung-Wook Min\altaffilmark{4},
Kwangsun Ryu\altaffilmark{4},
and Il-Joong Kim\altaffilmark{4},
}%
\affil{$^1$Korea Astronomy and Space Science Institute, 61-1 Hwaam-dong Yuseong-gu Daejeon, 305-348, Korea}
\affil{$^2$Space Sciences Lab., University of California, 7 Gauss Way, Berkeley, CA 94720-7450, USA}
\affil{$^3$Johns Hopkins University, Department of Physics and Astronomy, 3400 North Charles Street,
       Baltimore, MD 21218, USA}
\affil{$^4$Korea Advanced Institute of Science and Technology, 373-1 Guseong-dong, Yuseong-gu, Daejeon 305-701, Korea}
\email{kiseon@kasi.re.kr}

\begin{abstract}
We present far-ultraviolet (FUV) spectral images,
measured at \ion{C}{4} $\lambda$1550,
\ion{He}{2} $\lambda$1640, \ion{Si}{4}+\ion{O}{4}] $\lambda$1400, and \ion{O}{3}] $\lambda$1664,
of the entire Cygnus Loop,
observed with the Spectroscopy of Plasma Evolution from Astrophysical Radiation ({\em SPEAR})
instrument, also known as {\em FIMS}.
The spatial distribution of FUV emission generally corresponds with a limb-brightened
shell, and is similar to optical, radio and X-ray images.
The features found in the present work include a ``carrot'', diffuse interior,
and breakout features, which have not been seen in previous FUV studies.
Shock velocities of 140--160 km s$^{-1}$ is found from a line ratio of
\ion{O}{4}] to \ion{O}{3}], which is insensitive not only to resonance scattering
but also to elemental abundance.
The estimated velocity indicates that the fast shocks are widespread across the
remnant.
By comparing various line ratios with steady-state shock models,
it is also shown that the resonance scattering is widespread.
\end{abstract}%

\keywords{supernova remnants --- ultraviolet: ISM --- ISM: individual
(\objectname{Cygnus Loop}) --- shock waves}%

\section{Introduction}

The Cygnus Loop is one of the most well-studied supernova remnants (SNRs)
in our galaxy because of its large apparent angular size, its high surface brightness,
and its low reddening.
It is generally considered the prototypical ``middle-aged'' SNR.
Global features of the Cygnus Loop have been studied
in great detail at optical, X-ray, radio, and infrared wavelengths
\citep{Levenson1998, AL1999, LRB1997, ADL1992}.
X-rays are emitted from hot (temperature $T\sim10^6$ K) gas heated by shocks with
velocities $v_s \sim400$ km s$^{-1}$, while
optical emission arises from slower shocks
in which the postshock region cools to $T\sim10^4$ K.
Observations of far-ultraviolet (FUV) emission from
intermediate temperatures gas,
mostly limited to small spatial regions,
have played an important role for the
reliable estimation of shock speeds and elemental abundances.
Early studies concentrated on bright optical filaments
and compared FUV spectra with steady-flow shock models
\citep{Benvenuti1980, Raymond1980, Raymond1981, Raymond2001, Blair1991}.
Subsequent investigations, focused on fainter filaments
\citep{Raymond1983,Long1992,Sankrit2000,Sankrit2002},
have shown the complexities involved including the effects of
geometry, shock incompleteness, resonance line scattering,
dust grain destruction.

FUV observations of the entire Cygnus Loop are of great interest because
they provide a global understanding of the interaction between the X-ray
and optical emitting components.
\cite{CP1976} obtained images from a portion of the Cygnus Loop in two
broad bands (1250--1600 and 1050--1600\AA).
\cite{Blair1991} and \cite{RM1992} have produced
skymaps with nearly complete coverage of the Cygnus Loop,
measured at \ion{O}{6} $\lambda\lambda$1032, 1038.
Observations of five 40$\arcmin$ diameter fields
around the Cygnus Loop rim
with higher spatial resolution were obtained with the
UIT \citep{Cornett1992, Danforth2000}.

We report the first FUV spectral line images of the entire Cygnus Loop,
These images,
with spatial resolution of 3$\arcmin$ to 30$\arcmin$,
were obtained with {\em SPEAR} (The Spectroscopy of
Plasma Emission from Astrophysical Radiation), also known as {\em FIMS}
(Far-ultraviolet IMaging Spectrograph).
We describe how these data reveal physical conditions in the remnant and
complement earlier observations.

\section{Observations and Analysis}
{\em SPEAR} is a dual-channel FUV imaging spectrograph (Short channel `S': 900--1150\AA,
Long channel `L': 1330--1720\AA, $\lambda/\Delta\lambda\sim 550$)
with a large field of view (S channel: 4.0$^\circ\times4.6'$, L channel: 7.5$^\circ\times4.3'$),
designed to observe diffuse FUV emission lines.
{\em SPEAR} is the primary payload on the first Korean Science and Technology SATellite
(STSAT-1) and was launched into a $\sim$700 km sun-synchronous orbit on 27 September 2003.
The {\em SPEAR} instruments, their on-orbit performance, and the basic
processing of instrument data are described in detail by \citet{Edelstein2005a,Edelstein2005b}.

The Cygnus Loop was observed during 30 orbits between July 12 and July 22, 2004.
Of these, we used 11 orbits that include reliable attitude knowledge
of $\leq$ 30$\arcmin$.
The observation, totaling 6067 seconds, were taken in the 10\%
slit transmission mode.
A total of $\sim$1.6$\times$10$^5$ events were recorded.
The Cygnus Loop was observed by scanning the region along
the $\sim$4.5$\arcmin$ directions of the {\em SPEAR} field of view
with the field oriented nearly parallel to the $\alpha$ sky coordinate.
The aspect for these data was determined by comparing the positions of
observed bright stars with the TD-1 catalog \citep{Thompson1978}.
The position errors of the bright stars
along the scanning ($\Delta\delta$)
and spatial direction ($\Delta\alpha$) directions
were considered as
functions of $\delta$, ignoring errors in position angle.
The corrected positions of the bright stars coincided
to within 5$\arcmin$ of their TD-1 positions.
Position errors of all photons were interpolated linearly
from the bright stars' position errors.
All the photon events near the locations of the 11 brightest
TD-1 stars were removed to avoid stellar contamination.
The removed star pixels are indicated in Figure \ref{fig1}, together with
the names of characteristic SNR features.
Exposure maps were created by accumulating time-marked exposure event records to the sky
\citep{Edelstein2005b}.
Although the \ion{O}{6} emission line
from the Cygnus remnant is apparent in a preliminary analysis,
we do not further discuss the S channel data here
because difficult statistics require more extensive analysis.

To generate a net emission-line image,
we fit the spectrum in each pixel with
a constant continuum plus a spectral resolution-width
Gaussian function, fixed at the corresponding emission line center.
The approach,
basically an application of
a Bayesian method to estimate signal amplitude in the presence
of background assuming Poisson statistics \citep{Sivia1996},
avoids over-subtraction
that can occur by simply using line-adjacent spectral regions
for continuum removal.
The {\em SPEAR} Cygnus Loop images  obtained are sparsely occupied --
pixels may contain only a few counts.
To better estimate the surface brightness distribution,
we rebinned the images using a Gaussian function
with a position-variable angular scale
through an adaptive kernel method similar to that of \cite{HS1996} and \cite{EWR2000}.
The method overcomes fixed-size rebinning limitations that occur
when surface brightness or noise vary significantly across the image.

\section{Results}

The {\em SPEAR} FUV spectra from the entire Cygnus Loop SNR region
and from the characteristic features indicated in Figure \ref{fig1} are shown
in Figure \ref{fig2}.
A background spectrum has been subtracted using the data obtained from nearby faint skies.
The spectra clearly shows lines of many different ions including \ion{C}{4},
\ion{He}{2}, and \ion{N}{4}], and
the unresolved line complexes from
\ion{Si}{4}+\ion{O}{4}] and \ion{O}{3}].
It is also evident that the absolute and relative intensity of emission lines
vary between the regions.
Table \ref{table1} shows the reddening corrected FUV line luminosities for each region
as well as those averaged over the entire SNR.
A distance of 440 pc is assumed for their estimations \citep{Blair1999}.
The \ion{C}{3}, \ion{O}{6}, and X-ray luminosities, which are scaled
to the distance of 440 pc from the originally published values
\citep{Blair1991,Ku1984}, are also shown for comparison.
The \ion{C}{4} luminosity is a bit larger that the total 0.1--4 keV
X-ray luminosity.

Figure \ref{fig3} shows the spectral images of the entire Cygnus Loop obtained with {\em SPEAR}
for the \ion{C}{4} ($\lambda\lambda$1548, 1550\AA),
\ion{O}{3}] ($\lambda\lambda$1661, 1667\AA), \ion{He}{2} ($\lambda$1640\AA)
and \ion{Si}{4}+\ion{O}{4}] ($\lambda\lambda$1393, 1404\AA, unresolved).
The 0.03$^\circ$sky coordinates ($\alpha$, $\delta$) were binned with
smoothing kernel scales of $\sim$3$\arcmin$ to $\sim$30$\arcmin$ (1 $\sigma$ width).
To derive the \ion{C}{4} spectral image intensity, we have assumed 2:1 line ratio for
1548\AA\ and 1550\AA\ doublet lines applicable to a thin plasma.
For the \ion{Si}{4}+\ion{O}{4}] spectral image intensity, we
assumed two Gaussian lines having a $\sim$1:5.5 line ratio centered at 1393 and 1404\AA.
These line ratios were selected by averaging line ratios observed with {\em HUT}
\citep{Blair2002}.
Similarly, the \ion{O}{3}] lines are assumed to be formed in a doublet at $\lambda$1661
and $\lambda$1666 in the ratio of their statistical weights (0.41:1).
Signal to noise ratios of line intensities enclosed by the contours shown
in Figure \ref{fig3} are higher than at least two.
Although all of the FUV emission line images are generally similar,
differences in their detailed features are indeed apparent
as discussed below.

We compare the FUV spectra with
models \citep{HRH1987} that estimate
spectral line ratios as functions of shock velocity.
The models, while generated to predict parameters for regions of
higher density than encountered here, produce the correct relative
line intensities for the Cygnus region \citep{Cornett1992,
Blair1991}.
The model results can also be
complicated by many
effects, such as superposition of features, multiple shock velocities, and
resonance scattering \citep{Raymond1981,Danforth2000}.
Shock velocities for each image pixel were calculated from
various FUV line ratios
that were dereddened using $E(B-V)=0.08$ \citep{Miller1974}
and the average extinction curve of \citet{Cardelli1989}.
The resulting shock velocities were found to be 105--130, 90--100, and 140--180 km s$^{-1}$,
from the line ratios of \ion{C}{4}
to \ion{O}{3}], \ion{C}{4} to \ion{He}{2}, and \ion{O}{4}] to \ion{O}{3}],
respectively.
Here, the line intensity of \ion{O}{4}] was calculated from the intensity of
\ion{Si}{4}+\ion{O}{4}] assuming their line ratio of 1:5.5.
In most parts of the SNR, the
\ion{O}{4}]/\ion{O}{3}] estimated shock velocity
was in the range of 140--160 km s$^{-1}$.
Relatively uniform shock velocity distributions
were obtained except in the brightest position of the XA region
where an estimated velocity of 180 km s$^{-1}$ was found
from the \ion{O}{4}]/\ion{O}{3}] ratio.
\citet{Danforth2001} found substantially different line ratios
of \ion{Si}{4} to \ion{O}{4} at the XA region from the values
of \citet{Blair2002}.
Adopting the line ratio at the XA region,
we found no such high velocity region.

\section{Discussions}

\cite{Cornett1992} observed a 40$'$ diameter field named the ``XA'' region \citep{HC1986}
and centered at $\alpha\sim20^{\rm h}57^{\rm m}30^{\rm s}$,
$\delta\sim31^\circ7\arcmin36\arcsec$ with UIT.
They found the UIT feature closely follows a ``bow shock''
shape in {\em Einsten} HRI X-ray imagery.
The contours in Figure \ref{fig3} clearly show that the ``XA'' region is
the brightest region of the Cygnus Loop in the FUV emissions considered here,
although
the brightest positions in the FUV line maps do not exactly coincide.
Indeed, the ``XA'' region is seen to be brightest in many wavelengths.
\citet{HC1986} suggested that the enhanced ``XA'' region X-ray emission
is caused by a reflected shock resulting from a collision of the
blast wave with a cloud such as that seen in \ion{H}{1} maps of the area.

\begin{figure}
\begin{center}
\includegraphics[scale=.5]{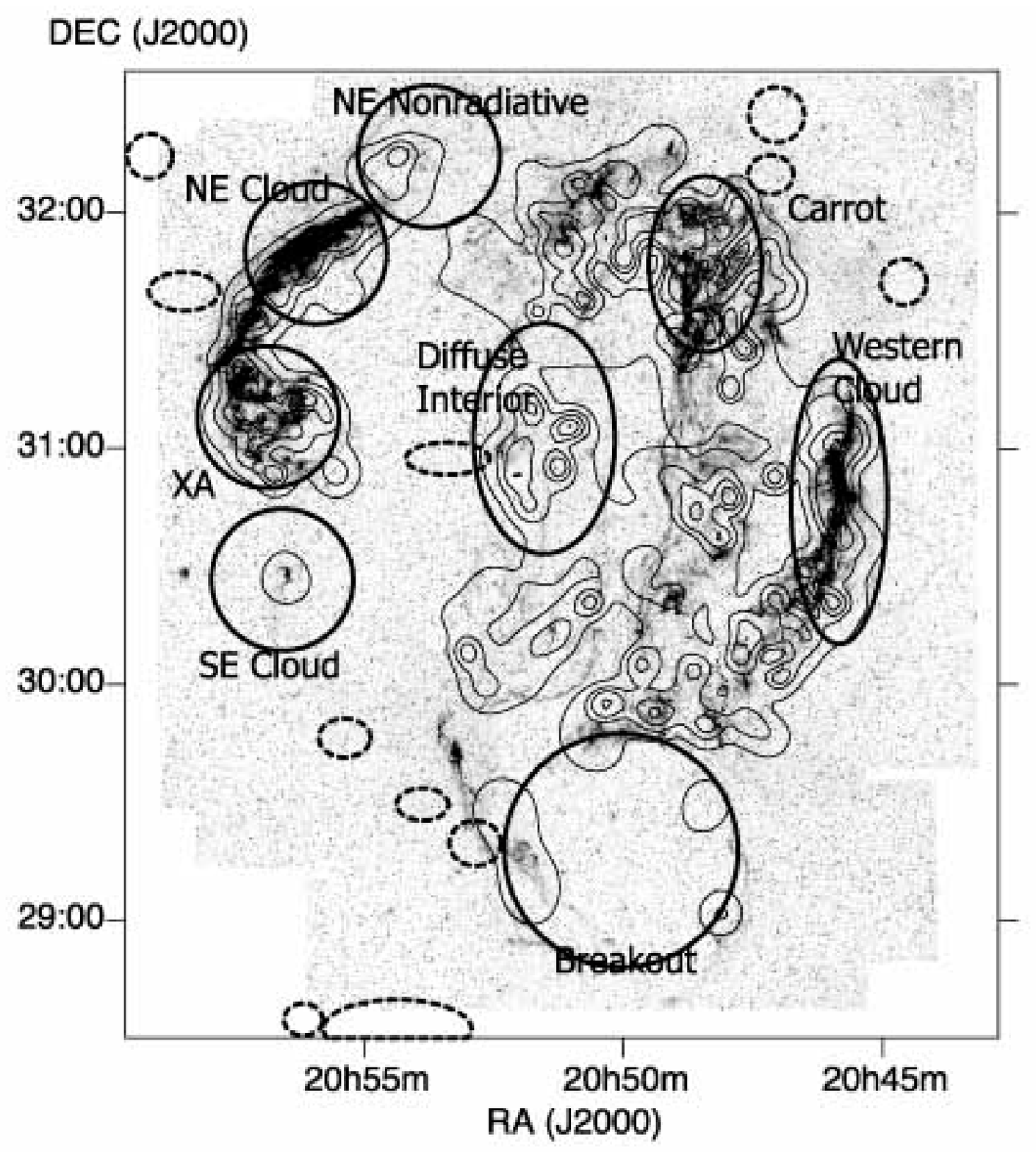}
\end{center}
\caption{
Contours of \ion{C}{4} ($\lambda\sim$ 1550\AA) emission intensity
are displayed over an H$\alpha$ image \citep{Levenson1998}
of the Cygnus Loop.
Thick solid lines identify characteristic features referred in the text.
Positions of the removed bright stars are indicated by dash lines.
} \label{fig1}
\end{figure}

\begin{figure}
\begin{center}
\includegraphics[scale=.4]{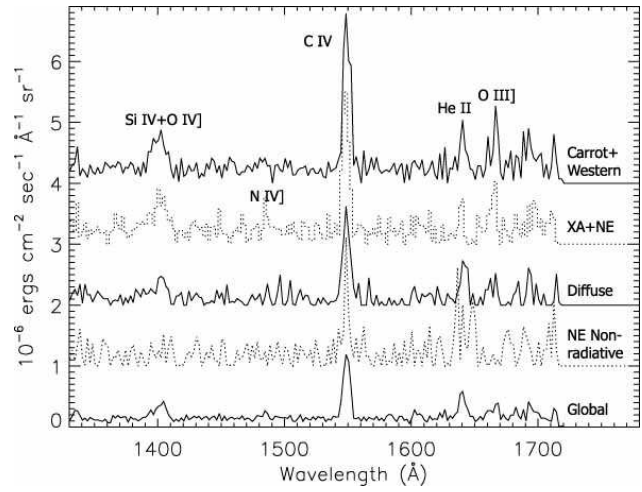}
\end{center}
\caption{The {\em SPEAR} L-channel spectra. Spectra of regions XA+NE cloud, Carrot+Wester cloud,
Diffuse interior, NE nonradiative are shown offset from each other.
Globally averaged spectrum is also shown.
The spectra are corrected for reddening.} \label{fig2}
\end{figure}

\begin{figure}
\begin{center}
\includegraphics[scale=.9]{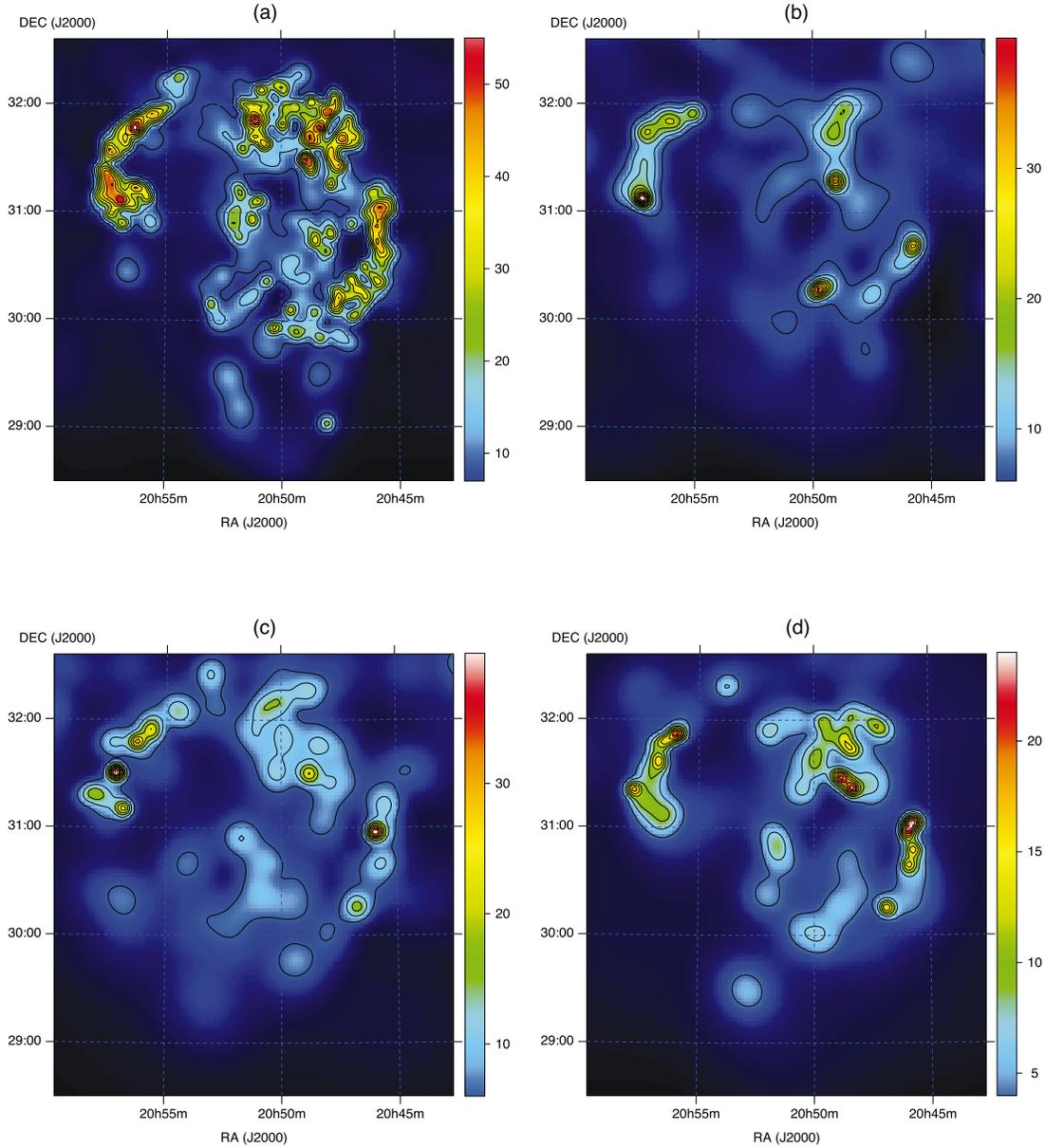}
\end{center}
\caption{{\em SPEAR} spectral line maps of the Cygnus Loop
in (a) \ion{C}{4} ($\lambda\lambda$ 1548, 1550\AA),
(b) \ion{O}{3}] ($\lambda\lambda$ 1661, 1667\AA), (c) \ion{He}{2} ($\lambda$ 1640\AA),
and (d) \ion{Si}{4}+\ion{O}{4}] ($\lambda\lambda$ 1400, 1403\AA, unresolved).
Coordinates are $\alpha , \delta$ of epoch 2000.0.
Units of color bars are 10$^{-6}$ ergs cm$^{-2}$ s$^{-1}$ sr$^{-1}$.
Contour levels in units of 10$^{-6}$ ergs cm$^{-2}$ s$^{-1}$ sr$^{-1}$ are stepped by
8 equal linear intervals, and range (a) from 7 to 55,
(b) from 6 to 40, (c) from 6 to 40, and (d) from 4 to 24.
The values are not corrected for reddening.
}
\label{fig3}
\end{figure}

The northeastern (NE) cloud radiative filaments, north of the XA region,
are clearly detected in the four FUV emission maps,
although again with differing locations of maximum brightness
which must indicates some range of morphological conditions.
The \ion{H}{1} associated with the NE cloud region has been identified
by \citet{Leahy2003}.
The \ion{H}{1} is a clear evidence for the clouds that the Cygnus Loop
shocks are interacting with.
The NE cloud extends into the southern portions of the northeastern
nonradiative field where
filaments extend counterclockwise from the northern limb and can
be seen prominently in H$\alpha$ \citep[e.g.,][]{Levenson1998}.
Portions of these filaments have been extensively studied by
\citet{Raymond1983}, \citet{Blair1991}, and \citet{Long1992}.
\citet{Danforth2000} presented an FUV image of the nonradiative filaments
with the UIT B5, which they conclude is a mixture mainly of \ion{C}{4}
and two-photon emission.
We observe \ion{C}{4} and \ion{He}{2} emissions from
the nonradiative region, as visible in Figures \ref{fig2} and \ref{fig3}.
Comparison of the \ion{C}{4} map with the H$\alpha$ maps of \citet{Levenson1998}
(see Fig.~\ref{fig1}(a)) shows that the
\ion{C}{4} emission follows the overall features of the H$\alpha$ emission.
The reddening-corrected flux of \ion{C}{4} we observe correspond to a
surface brightness of $1.0-4.3\times10^{-5}$ ergs cm$^{-2}$ s$^{-1}$ sr$^{-1}$,
consistent with the value, $3.9\times10^{-5}$ ergs cm$^{-2}$ s$^{-1}$ sr$^{-1}$,
from the ``south shock''
nonradiative front
observed with {\em HST} STIS \citep{Sankrit2000},
and a somewhat less than the value, $6.3\times10^{-5}$ ergs cm$^{-2}$ s$^{-1}$ sr$^{-1}$,
obtained with the {\em HUT} \citep{Long1992}.
Our maximum \ion{C}{4} flux value is a factor of $\sim$3 below
that reported by \citet{Raymond1983} and observed in the ``bright shock''
with the STIS \citep{Sankrit2000}.
We attribute this variation to the observatories different
fields of view.
We also note the relative weakness or nondetection of \ion{Si}{4}+\ion{O}{4}]
and \ion{O}{3}] emissions in the nonradiative
regions, as can be seen in Figures \ref{fig1}(b) and \ref{fig1}(d).
Radiative filaments show lines of
many different ions while nonradiative filaments show only lines of the
highest-ionization species \citep{Long1992, Danforth2000}.
Thus, we might conclude that the \ion{C}{4} and \ion{He}{2} emissions observed here
originate primarily from nonradiatve shocks,
although the {\em SPEAR} spatial resolution does not allow
a clear separation between non-radiative and
radiative regions.

The interior of the Cygnus Loop contains diffuse emission and filaments.
One noticeable interior feature in the FUV images (Fig.~\ref{fig2}) is the ``carrot''
(at $\alpha\sim20^{\rm h}49^{\rm m}$, $\delta\sim31{^\circ}40'$),
seen as a somewhat isolated vertical group
of filaments in the \ion{O}{3}] image.
A bright spot in the ``carrot''
(at $\alpha\sim20^{\rm h}49^{\rm m}$, $\delta\sim31{^\circ}20'$) coincides
in all four FUV spectral images.
Another region of diffuse emission in the \ion{C}{4}, \ion{He}{2}, and \ion{Si}{4}+\ion{O}{4}]
maps, also visible in optical images, runs
north-south across the center of the remnant,
around $\alpha\sim20^{\rm h}51^{\rm m}-52^{\rm m}$, $\delta\sim30^\circ20'-31^\circ10'$.

The western filaments were also detected in the four FUV emission lines, for which
the spatial features generally coincide.
\citet{Ku1984} found an X-ray bow-shaped shock front or a ``V'' feature on the southwest limb,
which has no optical counterparts.
\citet{AL1999} noted that the ``V'' feature has a cool right wing, and
a hot left wing.
A 1420-MHz image \citep{LRB1997} shows radio continuum emission
coincident with the ``V'' feature
that is particularly bright along the eastern side.
The \ion{C}{4} image in Figure \ref{fig1}(a) shows no counterparts to the ``V'' feature,
although the images show a bright feature to the north of the ``V''.
The absence of optical and FUV counterparts implies that the X-ray and
radio emission do not originate from dense radiative cooling regions but from
hotter gas, as suggested by \citet{Leahy2004}.

The breakout is the most significant departure from circularity in the remnant.
\citet{Leahy2002} discusses \ion{H}{1} observations of the Cygnus Loop which
show two cavities for the northern main part and the southern `breakout' part,
which are at two different velocities, and suggests that the cause may be either
one supernova explosion interacting with two blister regions
or two separate explosions.
The \ion{C}{4} image (Fig.~\ref{fig3}) shows the south breakout region,
particularly in its eastern portion.
The western region of the breakout is also weakly detected in the \ion{C}{4} map.
Figure \ref{fig1}(d) also shows marginal detection of the eastern region of
the breakout in \ion{Si}{4}+\ion{O}{4}] emission.

We also note that the southeastern (SE) cloud was detected in \ion{C}{4} and \ion{He}{2} emissions.
The region is an example of a cloud that is extended along the line of sight
but is not necessarily large across the plane of the sky \citet{Levenson1998}.
The \ion{H}{1} cloud associated with the SE region has been identified by \citet{Leahy2005}.
The boundary of the southeast corner and brightest part of the \ion{H}{1} ring matches
the boundary of the X-ray emission and of the optical and FUV emissions of the southeast knot.

As column density increases,
the \ion{C}{4} and \ion{He}{2} lines saturate due to resonance scattering effects,
while the intermediate lines,
such as \ion{O}{4}] and \ion{O}{3}], show little attenuation.
\cite{Cornett1992} and \cite{Danforth2000}
found that
resonance scattering in the strong FUV permitted lines is widespread in
the Cygnus Loop, especially in the bright optical filaments.
In this context, the line ratio \ion{O}{4}]/\ion{O}{3}] is an ideal
diagnostic for shock velocities.
The ratio is also insensitive to elemental abundance, as both lines
arise from the same element.
The ubiquity of \ion{O}{6} emission found by \citet{Blair1991}
implies that shocks with velocity greater than 160 km s$^{-1}$ are widespread
throughout the SNR.
The shock velocity estimated, using the ratio \ion{O}{4}]/\ion{O}{3}],
confirms that fast shocks are widespread in the Cygnus Loop.
In fact, the \ion{Si}{4} to \ion{O}{4}] ratio, used in the derivation of
\ion{O}{4}] line intensity, varies with shock velocity
\citep{Sankrit2003}.
Using the ratio $\sim$ 1:2 at shock velocity of 120 km s$^{-1}$, we obtained the
same shock velocity range.
\citet{Danforth2001} found, from a detailed analysis of
the XA region, that a significant column depth is present at all positions, including
those not near bright optical or UV filaments.
The lower velocity values obtained here from the ratio \ion{C}{4}/\ion{O}{3}]
and (even lower from) \ion{C}{4}/\ion{He}{2},
together with relatively uniform velocity-distributions,
also indicate that the resonance scattering in FUV permitted lines
is widespread across the Cygnus Loop.

\section{Concluding Remarks}

The {\em SPEAR} FUV spectral images of the Cygnus Loop offer a new
view of the SNR physical environment and conditions.
We find that the remnant's projected diffuse interior, ``carrot'', breakout region,
and periphery show correlated FUV, optical, X-ray, and radio emission.
The FUV emission can be well explained by the scenario wherein
the Cygnus Loop has been created by a cavity explosion \citep{Levenson1997},
with the presence of large clouds around its periphery \citep{HC1986, HRB1994, Leahy2002}.
Further study of these {\em SPEAR} data will provide insight into the overall
shock energetics through a more detailed
comparison with earlier observations.
These data will be particularly important for studying the NE nonradiative
shock,
where an additional set of {\em SPEAR} short-wavelength data
is expected to reveal large-scale features of \ion{O}{6} emission.

\acknowledgments
{\em SPEAR/FIMS} is a joint project of
Korea Astronomy and Space Science Institute,
Korea Advanced Institute of Science and Technology,
and the University of California at Berkeley,
funded by the Korea Ministry of Science and Technology
and the U.S. National Aeronautics and Space Administration Grant NAG5-5355.

\clearpage

\clearpage

\clearpage

\clearpage

\begin{table}
\caption{Reddening corrected FUV line luminosities of regions
indicated in Fig.~\ref{fig1}. The line luminosities are in units of $10^{34}$
ergs s$^{-1}$ and calculated assuming 440 pc distance \citep{Blair1999}.
The \ion{C}{3}, \ion{O}{6}, and X-ray luminosities from
\citet{Blair1991} and \citet{Ku1984} are also scaled to the distance
for comparison.}
\begin{tabular}{llllll}
\tableline
\tableline
Species & Carrot+Western & XA+NE & Diffuse & NE Nonradiative & Global \\ \tableline
\ion{Si}{4}+\ion{O}{4}] (1400\AA) & 0.67$\pm$0.18 & 0.52$\pm$0.16 & 0.26$\pm$0.10 &   & 6.6$\pm$0.6 \\
\ion{C}{4} (1550\AA) & 5.7$\pm$0.5 & 3.8$\pm$0.4 & 1.9$\pm$0.3 & 0.53$\pm$0.17 & 44.7$\pm$1.4 \\
\ion{He}{2} (1640\AA) & 0.54$\pm$0.17 & 0.21$\pm$0.12 & 0.34$\pm$0.12 & 0.15$\pm$0.09 & 6.8$\pm$0.6 \\
\ion{O}{3}] (1664\AA) & 0.88$\pm$0.24 & 0.55$\pm$0.20 & 0.35$\pm$0.14 & & 6.5$\pm$0.8 \\
\ion{C}{3} (980\AA) & & & & & 88.2 \\
\ion{O}{6} (1035\AA) & & & & & 150. \\
X-ray (0.1--4.0 keV) & & & & & 35.9 \\
\tableline
\end{tabular} \label{table1}
\end{table}

\end{document}